\documentclass[onecolumn]{article} % default is 10 pt
\usepackage{graphicx} % needed for including graphics e.g. EPS, PS
\usepackage{array}
\usepackage{multirow}
\usepackage{longtable}
\usepackage{amssymb}
\usepackage{amsmath}
\usepackage{amsthm}
\usepackage{framed}
\usepackage{a4wide}
\usepackage{enumerate}

\long\def\comment#1{}

\newcommand{\be}{\begin{equation}}
\newcommand{\ee}{\end{equation}}
\newcommand{\bea}{\begin{eqnarray}}
\newcommand{\eea}{\end{eqnarray}}
\newcommand{\beann}{\begin{eqnarray*}}
\newcommand{\eeann}{\end{eqnarray*}}
\newcommand{\bs}{\begin{split}}
\newcommand{\es}{\end{split}}

\newtheorem{thm}{Theorem}[section]
\theoremstyle{definition}

\theoremstyle{remark}

\theoremstyle{plain}
\newtheorem{lem}[thm]{Lemma}
 
\makeatletter
\def\@normalsize{\@setsize\normalsize{10pt}\xpt\@xpt
\abovedisplayskip 10pt plus2pt minus5pt\belowdisplayskip 
\abovedisplayskip \abovedisplayshortskip \z@ 
plus3pt\belowdisplayshortskip 6pt plus3pt 
minus3pt\let\@listi\@listI}

\def\subsize{\@setsize\subsize{12pt}\xipt\@xipt}
\def\section{\@startsection {section}{1}{\z@}{1.0ex plus
1ex minus .2ex}{.2ex plus .2ex}{\large\bf}}
\def\subsection{\@startsection 
   {subsection}{2}{\z@}{.2ex plus 1ex} {.2ex plus .2ex}{\subsize\bf}}
\makeatother

\begin{document}

% don't want date printed
\date{}

% >>>>>>>>>>>>>>>>>>>>>>>  Put your title here <<<<<<<<<<<<<<<<<<<<<<<<
% make title bold and 14 pt font (Latex default is non-bold, 16pt) 
\title{\Large {\bf A novel analytic operator method to solve linear ordinary differential equations with variable coefficients}}

% >>>>>>>>>>>>>>>>>>>>>>> Author's Name, Thanks or Affliation <<<<<<<<
\author{Wrick Sengupta \\{\it Department of Physics and Meteorology, IIT Kharagpur}\\Email: wricks@cts.iitkgp.ernet.in}

\maketitle
\thispagestyle{empty}

\subsection*{\centering Abstract}

% >>>>>>>>>>>>>>>>>>>>>>>>> Keywords and Abstract <<<<<<<<<<<<<<<<<<<<<
% Replace with your own keywords and abstract.  Text will be in italics
{\em
A new analytical operator method is discussed which solves linear ordinary differential equations with regular singularities. Solutions are obtained in analytic series form and also in Mellin-Barnes-type contour integral form. Exact series solution is obtained without having to calculate series coefficients by recurrence relation.
Both homogeneous and inhomogeneous equations are solved identically without having to calculate
the Green's function explicitly in the case of inhomogeneous equation.
Closed-form solutions are obtained for all the special functions appearing in
mathematical physics. For a second-order equation both the independent solutions are
obtained without invoking Wronskians, even when the indices differ by an integer.

{\bf Keywords:} 
special functions,exact solutions,contour integral solution,analytic series solution
}

\section{Introduction}

Operator methods for obtaining solutions of ordinary differential equations with 
constant coefficients are well known and well studied. However, there are very few 
operator methods in the literature to solve linear ordinary differential equations of all orders with variable coefficients.The normal method to solve homogeneous equations with regular singularities
is the Frobenius Method, where we assume a series and put it in the equation to obtain 
a recurrence relation between the series coefficients. However we face difficulties when the indices
at the regular singularities differ by an integer. The method involving Wronskians generally applied to
such cases is quite complicated. For inhomogeneous equations, calculation of Green's function using the solutions of the corresponding homogeneous equation is also difficult. There are really very few methods which treat homogeneous and
inhomogeneous equations in a simple and unified way.

 To provide such a unified method to solve ordinary and partial, homogeneous and inhomogeneous, linear
and nonlinear differential equations, the Adomian decomposition method \cite{adomian}  was designed. Using the inverse of the highest derivative, the differential equation is converted into an integro-differential equation, which is then solved by a successive approximation-type method to give an analytic series solution. While this method works beautifully in many cases, it fails severely when regular singularities are encountered. Out of a few thousands of papers published 
on the convergence and verification and modification of Adomian's method, very few really 
acknowledge this fact. One such is \cite{pdita}, where a modification of the invertible operator is proposed which takes care of the singularities. However, the authors have not given any particular way to solve the problem and have suggested that the form of the equation should decide the method. As an example they show how a lower-order operator (contrary to the inverse of the highest derivative in Adomian's method) can be used to obtain Bessel polynomials from the defining second-order differential equation, which has an irregular singularity. Though the authors have been successful in obtaining one solution ,there is no way to obtain the other one using their method. Also, no justification has been provided for using Picard's iteration on an integro-differential equation with a singular kernel.

 In this paper we shall give a general method to obtain solutions of ordinary linear differential equations having regular singularities. Using a particular kind of operator the differential equation shall be converted into an inhomogeneous integro-differential equation, the inhomogeneous part being just a function of initial conditions and the original inhomogeneous function, if present from before. The resolvent of the integro-differential operator acting on the inhomogeneous part gives the solution. 
 
 The integro-differential operator shall be shown to be bounded and contractive so that the well known Banach theory
can be used to obtain a binomial series expansion of the resolvent operator. The infinite series of the
operator then acts on the inhomogeneous part to generate the series solution. 
 
 In the next section we shall use a contour integral representation of the resolvent operator to 
 obtain Mellin-Barnes-type contour integral solutions to differential equations.

\section{Operator method to solve linear ordinary differential equations}
 
 The operator method can be used to solve general nth-order linear differential equations. We shall work mainly
with second-order ODEs in this paper and cover the most important special functions of mathematical physics which satisfy second-order ODEs. However the methods can be easily generalised to solve higher-order ODEs. We first study equations with a regular singularity at the origin and another (regular or irregular) singularity at infinity.

\subsection{Two singularities at $0$ and $\infty$}
The general second-order equation with singularities at $0$ and $\infty$ is:

 \begin{equation}
 {d^2\psi(z) \over{dz^2}}+p(z){d\psi(z) \over{dz}}+q(z)\psi(z) =F(z).
\end{equation}

If $p(z)$ and $q(z)$ have series expansions $p(z)=\sum_{i=-1}^{\infty } {p_i z ^i}$ and $q(z)=\sum_{i=-2}^{\infty } {q_i z ^i}$,
then the singularity at $z=0$ is regular \cite{morse}. We take a transformation $\psi (z)=z^\lambda f(z)$.
Such a transformation enables us to remove the term $z^{-2}$ from the coefficient of $f(z)$. Equating the coefficient of $z^{-2}$ to zero, we get the indicial equation
\be
\lambda^2+(p_{-1}-1)\lambda+q_{-2}=0.
\ee
 Let $\lambda_1$, $\lambda_2$ be the roots of the indicial equation, with $\lambda_1>\lambda_2$. If $\lambda_1-\lambda_2$ is an integer then the second solution would have a branch-point singularity or logarithmic singularity.
 The equation for a particular index $\lambda$, satisfied by $f(z)$, is
 
\begin{equation}
 {d^2 f(z) \over{dz^2}}+({\alpha \over z}+\sum_{i=0}^{n} {C_i z ^i}){df(z) \over{dz}}+(\sum_{i=0}^{n} {D_i z ^i})f(z) =z^{-\lambda}F(z),
\end{equation}

where $\alpha=2\lambda+p_{-1}$,$C_i=p_i$ and $D_i=\lambda p_i+q_{i-1}$. The above equation can be written as 

\begin{equation}
 {{1\over{z^\alpha }}{d\over{dz}}( z^\alpha {df(z) \over{dz}})}
+((\sum_{i=0}^{n} {C_i z ^i}){df(z) \over{dz}}+(\sum_{i=0}^{n} {D_i z ^i})f(z) )=z^{-\lambda}F(z).
\label{odefz}
\end{equation}

We define a new operator 
\be
L=\int{{1\over{z^\alpha }}(\int{z^\alpha(\cdot)dz})dz},
\label{def_L}
\ee

where the integrals are indefinite.
Operating  $L$  on (\ref{odefz}), we get

 \begin{equation}
f(z)+ \int{{1\over{z^\alpha }}(\int{z^\alpha((\sum_{i=0}^{n} {C_i z ^i}){df(z) \over{dz}}+(\sum_{i=0}^{n} {D_i z ^i})f(z))dz})dz}=\int{{1\over{z^\alpha }}(\int{z^{\alpha-\lambda}F(z)dz})dz}+f_0(z).
\label{odefzl}
\end{equation}

$f_0(z)$ is the 'integration constant' of the operator $L$, satisfying the constraint 
\be
{{1\over{z^\alpha }}{d\over{dz}}( z^\alpha {df_0(z) \over{dz}})}=0.
\label{odef0z}
\ee

From (\ref{odef0z}), we get 
\be
f_0(z)=c_0+c_1 \int{\frac{1}{z^\alpha}}dz.
\label{f0z}
\ee

We now define another operator
\be
A = \int{{1\over{z^\alpha }}
  (\int{z^\alpha((\sum_{i=0}^{n} {C_i z ^i}){d(\cdot) \over{dz}}+(\sum_{i=0}^{n} {D_i z ^i})(\cdot))dz})dz}.
\label{defA}
\ee

Thus, (\ref{odefzl}) can be written in operator notation as

\be
(1+A)\cdot f(z)= L\cdot z^{-\lambda }F(z)+f_0(z).
\label{op_eq}
\ee

The solution $f(z)$ of (\ref{op_eq}) can be formally written as 

\be
f(z)=(1+A)^{-1}\cdot L\cdot z^{-\lambda }F(z) + (1+A)^{-1}\cdot f_0(z).
\label{op_eqsoln}
\ee

From the above equation, we can see that first term on the right-hand side is the particular solution while
 the second part is the complementary solution, which is also the solution of the corresponding 
homogeneous equation. The Green's function operator can be seen to be
$(1+A)^{-1}\cdot L\cdot z^{-\lambda }()$.

\subsection{Three singularities at $0$, $1$ and $\infty$}
The general second-order equation with singularities at $0$, $1$ and $\infty$ can be written as:

 \begin{equation}
 z(1-z){d^2\psi(z) \over{dz^2}}+p(z){d\psi(z) \over{dz}}+q(z)\psi(z) =F(z).
\end{equation}

If $p(z)$ and $q(z)$ have series expansions $p(z)= z \sum_{i=-1}^{\infty } {p_i z ^i}$ and $q(z)= z \sum_{i=-2}^{\infty } {q_i z ^i}$, then we have a regular singularity at $0$. If the differential equation has singularities at any 3 other points, then the singularities can be mapped onto $0$, $1$ and $\infty$ without changing the indices, as explained in \cite{morse}.

As before, we take a transformation $\psi (z)=z^\lambda f(z)$ to remove the term $z^{-2}$ from the coefficient of $f(z)$. We obtain the indices by solving the indicial equation likewise.
 The equation for a particular index $\lambda$, satisfied by $f(z)$, is
 
\begin{equation}
 {d^2 f(z) \over{dz^2}}+({\alpha \over z}+\sum_{i=0}^{n} {C_i z ^i}){df(z) \over{dz}}+(\sum_{i=0}^{n} {D_i z ^i})f(z)- z{d^2 f(z) \over{dz^2}} =z^{-\lambda}F(z),
\end{equation}

where $\alpha=2\lambda+p_{-1}$, $C_0=p_0-2\lambda$, $D_0=\lambda(1-\lambda)+\lambda p_0+q_{-1}$, $D_1=\lambda(1-\lambda)+\lambda p_1+q_{0}$ and $D_i=\lambda p_i+q_{i-1}$, $C_i=p_i$ for all other $i$. The above equation can be written as

\begin{equation}
 {{1\over{z^\alpha }}{d\over{dz}}( z^\alpha {df(z) \over{dz}})}
+((\sum_{i=0}^{n} {C_i z ^i}){df(z) \over{dz}}+(\sum_{i=0}^{n} {D_i z ^i})f(z)- z \frac{d^2 f(z)}{dz^2})=z^{-\lambda}F(z).
\label{3odefz}
\end{equation}

Defining $L$ as in (\ref{def_L}) and operating $L$ on (\ref{3odefz}), we get, as before,
\be
(1+A)\cdot f(z)= L\cdot z^{-\lambda } F(z)+f_0(z),
\label{3op_eq}
\ee

where $f_0(z)$ is as defined before in (\ref{f0z}), and 

\be
A = \int{{1\over{z^\alpha }}
  (\int{z^\alpha(-z \frac{d^2 (\cdot)}{dz^2}+( \sum_{i=0}^{n} {C_i z ^i}){d(\cdot) \over{dz}}+(\sum_{i=0}^{n} {D_i z ^i})(\cdot))dz})dz}.
\label{3defA}
\ee

The solution $f(z)$ can again be written as (\ref{op_eqsoln})

In the next section, we shall study the properties of the operators $L$ and $A$.

\section{The operators $L$ and $A$}

Let $G$ be an open subset of $\mathbb{C}$ defined by $\{z \in G:0<|z|\leq 1\}$. Let $v : G\rightarrow \mathbb{R_{+}}$ be a continuous and strictly positive function on $G$, here called the weight.
We define the following weighted Banach space of holomorphic functions on $G$:
\be
Hv(G) := {\{f \in H(G):||f||_{v} := \mathop{\rm \sup}\limits_{v\in G}\,v(z)|f(z)| <   \infty\}}
\label{def_hv}
\ee

We refer to \cite{galbis,point,comp,note}, and the references given therein, for details on the above Banach space.
 From the standard literature on solutions of second-order ODEs, it is obvious that our space contains functions of the regular and irregular kinds.
 The regular kind is finite at the origin and has a Taylor series expansion about the origin:
 \be
 f_r(z)=\sum_{i=0}^{\infty}cr_i z^i.
 \label{def_fr}
 \ee

 The irregular one has multiple pole and even branch-point singularity at the origin
\be
f_{ir}(z)=
\begin{cases}
 z^{-\delta  \lambda}\sum_{i=0}^{\infty}cn_i z^i 
& \text{if $(\lambda _1-\lambda _2)\not\in \mathbb{N}$},\\

(\log z) \sum_{i=0}^{\infty}c_i z^i
& \text{if $(\lambda _1=\lambda _2)$},\\

 z^{-\delta  \lambda}\sum_{i=0}^{\infty}cn_i z^i 
+ (\log z) \sum_{i=0}^{\infty}c_i z^i
 & \text{if $(\lambda _1-\lambda _2)\in \mathbb{N}$},
\end{cases}
\label{def_fir}
\ee
where $\delta  \lambda=\lambda _1-\lambda _2=\alpha -1 $.

We choose $v(z)=|z|^\alpha $, so that 
\be
\lim_{|z|\rightarrow 0} f(z)v(z)=0
\ee

for both regular and irregular functions.

\begin{thm}
 There is a subspace $G_0=\{z \in G_0:0<|z|\leq z_0\}\subset G$ such that\\
i.   $L : Hv(G_0)\rightarrow Hv(G_0)$ is continuous,\\
ii.  $A : Hv(G_0)\rightarrow Hv(G_0)$ is continuous, and\\
iii. $(1+A)^{-1}=1-A+A^2-A^3+A^4+...+(-1)^{j}A^j+...$
\end{thm}

\begin{proof}
i. For $\alpha\not\in \mathbb{N}$ we have $f(z)=\sum_{m=-{\delta  \lambda}}^{\infty} c_m z^m$.

From the definition of $L$ in (\ref{def_L}), we have:
$
 L z^{m}  ={\int{\frac{1}{z^\alpha }}(\int{z^{\alpha+m}dz})dz}=  \frac{z^{m+2}}{(m+2)(\alpha +m+1)}.
$
Therefore $||L z^{m}||_{v}\leq M ||z^m||_v$ where $ M=\frac{z^{2}}{(m+2)(\alpha +m+1)}\leq 1$.

For $\alpha\in \mathbb{N}$ we have integral powers of $z$ starting from $z^{1-\alpha}$ (for $\alpha\neq 1$) as well as $ z^{m}\log(z)$ terms in $f(z)$ with $m=0,1,2..\infty$.
For $\alpha=1$,

$ L \cdot z^{m}  =  \frac{z^{m+2}}{(m+2)^2}$ and $L z^{m}\log(z)  =  \frac{\log(x)}{(m+2)^2}-2 \frac{z^{m+2}}{(m+2)^3}.$

For $\alpha\in \mathbb{N}$ but $\alpha\neq 1$ and $m=0,1,2..\infty $,

\begin{equation*}
L z^{1-\alpha+m}=
  \begin{cases}
   \frac{z^{3+m-\alpha}}{(m+2)(3+m-\alpha)} & \text{if $m\neq \alpha-3$},\\
   \frac{\log(z)}{(\alpha-1)} & \text{if $m=\alpha-3$}.
  \end{cases}
\end{equation*}

For the $z^n \log(z)$ terms with $n\in \mathbb{R_+}$,

$$L\cdot z^{m}\log(z)=z^{2 + m} (\frac{-3 -2 m - \alpha}{(2 +m)^2 (1 + m + \alpha)^2} + \frac{\log(z)}{(2 +m) (1 + m+\alpha)}).$$

Thus, from all the above cases, we can see that for $f$ defined as in (\ref{def_fr}, \ref{def_fir}), $L:f\rightarrow f$. Also,
$||L f(z)||_{v}\leq ||f(z)||_v$ for all $0<|z|\leq |z_0|<1$.  
This proves that $L$ is continuous.

ii. For $A$ defined as in (\ref{defA}), we can immediately see that for $\alpha\not\in \mathbb{N}$,
\be
\nonumber A \cdot z^{m}=\sum_{i=0}^{\infty}\frac{(m+\lambda)p_i+q_{i-1}}{(i+m+\alpha)(i+m+1)}z^{i+m+1}.
\ee
The series obtained is seen to be convergent from the ratio test. Therefore, $||A z^{m}||_{v}\leq M ||z^m||_v$, where $ M=\sup \sum_{i=0}^{\infty}\frac{(m+\lambda)p_i+q_{i-1}}{(i+m+\alpha)(i+m+1)}z_{0}^{i+1}$. $f(z)=\sum_{m=-{\delta  \lambda}}^{\infty} c_m z^m$, therefore $||A f(z)||_{v}\leq M ||f(z)||_v$.

For the $z^n \log(z)$ terms with $n\in \mathbb{R_+}$,

$L\cdot z^{m}\log(z)=\sum_{i=0}^{\infty}\frac{-(\lambda p_i+q_{i-1})(2 i+ 2k+ \alpha+1)+i+i^2-k^2+\alpha(1+i)}{(i+m+\alpha)^2(i+m+1)^2}z^{i+m+1} + \sum_{i=0}^{\infty}\frac{(m+\lambda)p_i+q_{i-1}}{(i+m+\alpha)(i+m+1)}z^{i+m+1}\log(z)$.

For $A$ defined as in (\ref{3defA}), we have just another extra term $L(z \frac{d^2f(z)}{dz^2})$.

$L(z \frac{d^2(z^m))}{dz^2})=\frac{m(m-1)}{(m+1)(m+\alpha)}z^{m+1}$ for $m\neq-1$ and $m\geq1-\alpha$

 For  $m=-1$, $L(z \frac{d^2(z^{-1}))}{dz^2})=\frac{2}{1-\alpha}\log(z)$.

$L(z \frac{d^2(z^m\log(z)))}{dz^2})=\frac{(-\alpha + 2 m \alpha + m^2 (2 + \alpha) + m (-1 + m^2)(m + \alpha) \log(z))}{(m+1)^2(m+\alpha)^2}z^{m+1}$ for $m\neq-1$.

From the above, it is seen that $A:f\rightarrow f$ and $||L f(z)||_{v}\leq M ||f(z)||_v$.

iii. We can choose $z_0$ such that $M<1$. Therefore, for $0<|z|\leq |z_0|<1$, $A$ is a contraction operator and hence we can expand the resolvent $(1+A)^{-1}$ in a Neumann series.

\end{proof}

\section{Representations of the resolvent operator}

In the last section, we saw that the resolvent operator $(1+A)^{-1}$ can be expanded in a Neumann series for $0<|z|\leq |z_0|<1$, so that $M<1$. We can also have a contour-integral representation of $(1+A)^{-1}$:
\be
(1+A)^{-1}x(z)=\frac{1}{2 \pi \iota}\int_{a-\iota \infty}^{a+\iota \infty}\Gamma(s)\Gamma(1-s)(-A)^{-s}x(z)ds,
\label{def_resolvent}
\ee 
where $0<a<1$ and $x(z)$ is an element of the Banach space $Hv(G)$.

This form of the resolvent operator can be obtained by taking an inverse Mellin transformation of the expression for the fractional power of a closed operator given by V. Balakrishnan in \cite{balakrishnan} and \cite{fA}.

We consider the expression $\Phi(t)=\frac{1}{2\pi \iota}\int_{C}\Gamma(s)\Gamma(1-s)(-t)^{-s}ds$, where $C$ is a Mellin-Barnes-type contour. For $|t|<1$ we close the contour on the left side of the complex plane, and so the singularities enclosed by the contour are those of $\Gamma(s)$ at $s=-n$, $n\in\mathbb{Z+}$. Using Cauchy's residue theorem, we obtain $\Phi(t)=\sum_{i=0}^{\infty}t^{i}$. Similarly, the expression for the resolvent operator defined by (\ref{def_resolvent}) gives the Neumann series for $||A||_{v}<1$.

In the next section, we shall apply the operator method to obtain exact solutions of various well-known differential equations occurring in mathematical physics.

\section{Examples from Mathematical Physics}

\subsection*{A. Homogeneous equations}

\subsection{Singularities at $\infty$}
To begin with, we solve the most elementary differential equations with constant coefficients.
\subsubsection{The exponential function}

The first-order ODE satisfied is $\frac{dy}{dz}=y(z)$.We integrate both sides of the equation to get
\be
(1-\mathfrak{I})y(z)=c,
\ee

where $\mathfrak{I}$ is the indefinite integration operator and $c$ is the integration constant. Thus, our operator $A$ is $\mathfrak{I}$. For $|z|<1$, $||\mathfrak{I}||_{v}<1$. 
The solution is
\be
y(z)=(1-\mathfrak{I})^{-1}c.
\ee

We first expand the resolvent in a Neumann series. So,
\be
y(z)=(1+\mathfrak{I}+\mathfrak{I}^2+\mathfrak{I}^3...)c= c(1+z+\frac{z^2}{2!}+\frac{z^3}{3!}+...)=c e^z.
\ee

Next we use the contour-integral form of the resolvent. A fractional power of $\mathfrak{I}$ is identical to that of the definite integral with limits $0$ and $z$. We know from fractional calculus that 
\be
\mathfrak{I}^{-s}c=c \frac{z^{-s}}{\Gamma(1-s)}.
\label{Ionc}
\ee

Therefore,
\be
(1-\mathfrak{I})^{-1}c=\frac{1}{2\pi \iota}\int_{a-\iota \infty}^{a+\iota \infty}\Gamma(s)\Gamma(1-s)(-\mathfrak{I})^{-s}cds
=\frac{1}{2\pi \iota}c\int_{a-\iota \infty}^{a+\iota \infty}\Gamma(s)(-z)^{-s}ds.
\ee

The last expression is the inverse Mellin transform of gamma function, which is the exponential function.

\subsubsection{Trigonometric and hyperbolic functions}

The second-order differential equation satisfied by these are $\frac{d^2y}{dz^2}\pm  \omega^2 z=0$. We choose $L=\mathfrak{I}^2$. Acting both sides by $L$, we get 
\be
(1\pm \mathfrak{I}^2\omega^2)y(z)=c_0+c_1 z,
\ee
where $c_0 and c_1$ are constants.

Therefore $A$ is $\pm \mathfrak{I}^2\omega^2$. We choose $|z|<\omega^{-1}$ to ensure that $||A||_v<1$.
The Neumann series expansion method proceeds exactly as before. We get cos$(z)$ and cosh$(z)$ with $c_0$, and sin$(z)$ and sinh$(z)$ with $c_1$.

We have
\be
\mathfrak{I}^{p}z^q=c \frac{\Gamma(1+q)z^{q+p}}{\Gamma(1+q+p)}.
\label{Iponzq}
\ee

Substituting $p=-2s$, $q=0$ for the term with $c_0$, and $q=1$ for the term with $c_1$,  we get 

\begin{align}
\nonumber y(z)&=(1+A)^{-1}(c_0+c_1 z)\\
\nonumber&=\frac{1}{2\pi \iota}\int_{a-\iota \infty}^{a+\iota \infty}\Gamma(s)\Gamma(1-s)(\mp\omega^2 \mathfrak{I}^2)^{-s}(c_0+c_1 z)ds\\
&=\frac{1}{2\pi \iota}c_0\int_{a-\iota \infty}^{a+\iota \infty}\frac{\Gamma(s)\Gamma(1-s)}{\Gamma(1-2s)}(\pm \omega z)^{-2s}ds + \frac{1}{2\pi \iota}\frac{c_1}{\omega}\int_{a-\iota \infty}^{a+\iota \infty}\frac{\Gamma(s)\Gamma(1-s)}{\Gamma(1-2s)}(\pm \omega z)^{2-2s}ds.
\end{align}

The above contour integrals, though not the exact inverse Mellin transforms, can be shown to yield the same trigonometric / hyperbolic functions, using Cauchy's residue theorem.

\subsection{Singularities at $0$ and $\infty$}

We treat the most important special functions having singularities at $0$ and $\infty$ $-$ the Bessel functions and the confluent hypergeometric function.

\subsubsection{The Bessel functions}
\label{bessel}

\begin{equation}
 {d^2\psi(z) \over{dz^2}}+{1\over{z}}{d\psi(z) \over{dz}}+(1 -{\nu ^2\over{z^2}})\psi(z) =0.
\end{equation}

This equation has a regular singularity at $z=0$ and an irregular singularity at $\infty$.We take $\psi(z)=z^\kappa f(z)$. The indices at the regular singularity are $\kappa=\pm \nu $. The equation for $f(z)$ for the positive index is

 \begin{equation}
 {d^2f(z) \over{dz^2}}+{(2\nu  + 1)\over{z}}{df(z) \over{dz}}+ f(z) =0,
\end{equation}
which can be written as 
 \begin{equation}
 {\frac{1}{z^{2\nu +1}}{d\over{dz}}( z^{2\nu +1} {df(z) \over{dz}})} =- f(z).
\end{equation}
We choose the operator $L=\int\frac{1}{z^{2\nu +1} }(\int z^{2\nu +1}()dz)dz$. 

The `integration constant' $f_0(z)$ for $L$ satisfies 

\be
\frac{1}{z^{2\nu +1}}\frac{d}{dz}( z^{2\nu +1} \frac{df_0(z)}{dz})=0.
\ee

A simple integration of the above equation gives $f_0(z)=c_0+c_1\int\frac{dz}{z^{2\nu +1}}$.
 
Acting on both sides by $L$, we get $(1+A)f(z)=f_0(z)$, where $A= L$ is a bounded operator.We choose $z_0$ suitably, so that $\forall$ $|z|<|z_0|$, $||A||_v<1$.

The regular solution is obtained by acting the resolvent operator on $c_0$. We start with the Neumann series as usual.

$A \cdot c_0= c_0 \frac{1}{1+\nu}\ (\frac{z}{2})^2$, $A^2 \cdot c_0= c_0 \frac{1}{2(1+\nu)(2+\nu)}\ (\frac{z}{2})^4...$

It can be proved by induction that

$A^n \cdot c_0=\frac{1}{\prod_{j=1}^{n}(j)(j+\nu)}(\frac{z}{2})^{2n}$.

Therefore, the regular solution is
\begin{align}
\nonumber fr(z)&=(1-A+A^2+...)c_0\\
&=c_0\biggl(1-\frac{1}{(1+\nu)}\ \biggl(\frac{z}{2}\biggr)^2+\frac{1}{2(1+\nu)(2+\nu)}\ \biggl(\frac{z}{2}\biggr)^4+...\biggr),
\end{align}

which is just $c_0 \Gamma(1+\nu)(\frac{z}{2})^{-\nu}J_n(z)$.

Since the calculation of fractional powers of operators is difficult, a heuristic method to obtain the fractional power shall be followed from now on. The method is to calculate $A^n$ for $n \in \mathbb{N}$ and then to generalise the expression so that $n$ can be fractional too.

In this case, we see that writing  
\begin{align}
\nonumber A^n \cdot c_0&= c_0 \frac{1}{n! (1+\nu)_n}\biggl(\frac{z}{2}\biggr)^{2n}\\
&= c_0\frac{\Gamma(\nu+1)}{\Gamma(1+\nu+n)\Gamma(1+n)}\biggl(\frac{z}{2}\biggr)^{2n}
\label{anonc0bes}
\end{align}

solves our problem, as the $\Gamma(n)$ functions are valid for both integral and nonintegral $n$. Therefore,

\begin{align}
\nonumber fr(z)&= \frac{1}{2 \pi \iota}\int_{a-\iota \infty}^{a+\iota \infty}\Gamma(s)\Gamma(1-s)(-A)^{-s}c_0ds\\
&= \frac{1}{2 \pi \iota}c_0\int_{a-\iota \infty}^{a+\iota \infty}\frac{\Gamma(s)}{\Gamma(1+\nu-s)}\biggl(-\frac{z}{2}\biggr)^{-2s}ds.
\end{align}

When $\nu \notin \mathbb{R+}$, then the second irregular solution can be easily obtained by acting the resolvent on the term with $c_1$.

$A \cdot c_1\frac{z^{-2\nu}}{{-2\nu}}= \frac{c_1}{{-2\nu}} \frac{1}{1-\nu}\ (\frac{z}{2})^2$, $A^2 \cdot c_1\frac{z^{-2\nu}}{{-2\nu}}= \frac{c_1}{{-2\nu}} \frac{1}{2(1-\nu)(2-\nu)}\ (\frac{z}{2})^4... $

For general $v$,

\be
A^v\biggl(\int z^{1+2\nu}\biggr)=(-1)^v \frac{\Gamma(1-\nu)}{\Gamma(1-\nu+v)\Gamma(1+v)}\biggl(\frac{z}{2}\biggr)^{2v-2\nu}.
\label{amirbes}
\ee

Thus we obtain the irregular function as
\begin{align}
\nonumber fi(z)&=z^{-2\nu}\frac{c_1}{{-2\nu}}\biggl(1-\frac{1}{(1-\nu)}\ \biggl(\frac{z}{2}\biggr)^2+\frac{1}{2(1-\nu)(2-\nu)}\ \biggl(\frac{r}{2}\biggr)^4+...\biggr)\\
&=\frac{1}{2 \pi \iota}\frac{c_1}{{-2\nu}}\int_{a-\iota \infty}^{a+\iota \infty}ds\frac{\Gamma(s)}{\Gamma(1-\nu-s)}\biggl(-\frac{z}{2}\biggr)^{-2s-2\nu},
\end{align}
which is $\frac{c_1}{{-2\nu}} \Gamma(1-\nu)(\frac{z}{2})^{-\nu}J_{-n}(z)$.

In fact equations (\ref{anonc0bes},\ref{amirbes}) are special cases of

\be
A^v(x^q)=\frac{(-1)^v x^q}{(1+\frac{q}{2})_v(1+\frac{q}{2}+v)_v}\biggl(\frac{x}{2}\biggr)^{2v}
\label{anonxqbes}
\ee
where $(a)_v$ is the Pochhammer symbol.

When $\nu=n \in \mathbb{R_{+}}$, logarithmic terms occur. We need a result from fractional calculus:
\be
\mathfrak{I}^v(z^{\lambda}\ln(z))=\frac{\Gamma(\lambda+1)}{\Gamma(\lambda+v+1)}z^{\lambda+v}(\ln(z)+\psi(\lambda+1)-\psi(\lambda+v+1)),
\ee
where $\psi$ is the digamma function with $\psi(1)=-\gamma$, the Euler constant.

We need to obtain $A^{-s} \int z^{1+2n}$. We observe that equation (\ref{amirbes}) is valid for $\nu=n$ with $v \leq n-1$. 

For $v=n$, we have 
\be
A^n(\int z^{1+2n})=\frac{(-1)^n\ln(z)}{(2^n\Gamma(1+n))^2}.
\label{m_eql_n}
\ee

$A^{n+1}(\int z^{1+2n})=(-1)^n\frac{1}{n+1}(\frac{z}{2})^2(\ln(z)-\frac{n+2}{2(n+1)})$,
$A^{m+n}(\int z^{1+2n})=(-1)^n(\frac{z}{2})^{2m}(c_{1}(m)\ln(z)+c_{2}(m))$ for integral m.

Therefore,
\begin{align}
\nonumber A^{m+n+1}\biggl(\int z^{1+2n}\biggr)=(-&1)^n\biggl(\frac{z}{2}\biggr)^{2(m+1)}\\
&\biggl(\frac{c_{1}(m)}{(1+m)(1+m+n)}\ln(z)+\frac{2(1+m)(1+m+n)c_{2}(m)-c_1(m)(2+2m+n)}{2(1+m)^2(1+m+n)^2} \biggr)
\end{align}

From the above, we obtain 
\begin{align}
c_1(m+1)&=\frac{c_{1}(m)}{(1+m)(1+m+n)},\label{c1}\\
c_2(m+1)&=\frac{2(1+m)(1+m+n)c_{2}(m)-c_1(m)(2+2m+n)}{2(1+m)^2(1+m+n)^2}.
\label{c2}
\end{align}

 We uniquely solve (\ref{c1},\ref{c2}) with the condition (\ref{m_eql_n}) to get 

\begin{align}
c_1(m)  =&(-4)^{-n} \frac{\Gamma(2 + n)}{(1 + n) \Gamma(1 + m) \Gamma(1 + n)^2 \Gamma( 1 + m + n)},\\
c_2(m)  =&-\frac{(-1)^n \Gamma(2 + n)}{2^{(1+2n)}(1 + n) \Gamma(1 + m) \Gamma(1 + n)^2 \Gamma(1 + m + n)}
\\
\nonumber&(\gamma + \psi(1 + m) - \psi(1 + n) + \psi(1+m+n)),\\
A^{m+n}\biggl(\int z^{1+2n}\biggr) =&\frac{(-1)^n(2\ln(z)-\gamma+\psi(n+1)-\psi(m+1)-\psi(m+n+1))}{2^{(1+2n)}\Gamma(n+1)\Gamma(m+1)\Gamma(n+m+1)}\biggl(\frac{z}{2}\biggr)^{2m}.
\label{amlnz}
\end{align}

 Using (\ref{amlnz}) and putting $v=m+n$, we get 
 \be
 A^{v}\biggl(\int z^{1+2n}\biggr) =\frac{(-1)^v(2\ln(z)-\gamma+\psi(n+1)-\psi(v-n+1)-\psi(v+1))}{2^{(1+2n)}\Gamma(n+1)\Gamma(v-n+1)\Gamma(v+1)}\biggl(\frac{z}{2}\biggr)^{2(v-n)}.
\label{avlnz}
 \ee
 
This equation is valid for fractional $v$ as well. The irregular solution for integral $n$ can thus be written as 
\begin{align}
\nonumber
fr(z) =& \frac{1}{2 \pi \iota}\int_{a-\iota \infty}^{a+\iota \infty}\Gamma(s)\Gamma(1-s)(-A)^{-s}\biggl(\int\frac{1}{z^{2n+1}}\biggr)ds\\
\nonumber
 =&\frac{1}{2 \pi \iota}\frac{2\ln(z)-\gamma+\psi(n+1)}{\Gamma(n+1)}\int_{a-\iota \infty}^{a+\iota \infty}\frac{\Gamma(s)}{2^{2n+1}\Gamma(1-s-n)}\biggl(\frac{z}{2}\biggr)^{-2(s+n)}ds \\
 &+\frac{1}{2 \pi \iota}\frac{1}{\Gamma(n+1)}\int_{a-\iota \infty}^{a+\iota \infty}\frac{\Gamma(s)(\psi(1-s-n)+\psi(1-s))}{2^{2n+1}\Gamma(1-s-n)}\biggl(\frac{z}{2}\biggr)^{-2(s+n)}ds
\end{align}

\subsubsection{The confluent hypergeometric function}

\begin{equation}
 \frac{d^2\psi(z)}{dz^2}+\biggl(\frac{c}{z}-1\biggr)\frac{d\psi(z)}{dz} -\frac{a}{z}\psi(z) =0.
\end{equation}

The indices are $0$ and $1-c$.  The ODE can be written as 
 \begin{equation}
 \frac{1}{z^{c}}\frac{df(z)}{dz}\biggl( z^{c} \frac{df(z)}{dz}\biggr) - \biggl(\frac{df(z)}{dz}+\frac{a}{z}f(z)\biggr)=0.
\end{equation}
We choose the operator $L=\int\frac{1}{z^{c} }(\int z^{c}()dz)dz$. 
The `integration constant' $f_0(z)$ for $L$, satisfying $\frac{1}{z^{c}}\frac{d}{dz}( z^{c} \frac{df_0(z)}{dz})=0$,
is $f_0(z)=c_0+\int{z^{-c}}dz$. Operating by $L$, we get $(1-A)f(z)=f_0(z)$, where $A=\int\frac{1}{z^{c}}(\int z^{c}((\frac{d()}{dz}+\frac{a}{z}()))dz)dz$.

\begin{lem}
For $q\in \mathbb{R_+}$ but $q\notin \mathbb{N}$,
\be
\nonumber A^m(z^q)=\frac{(q+a)_m}{(q+c)_m(q+1)_m}z^{q+m}=\frac{\Gamma(a+m+q)\Gamma(1+q)\Gamma(q+c)}{\Gamma(a+q)\Gamma(c+m+q)\Gamma(1+m+q)}z^{q+m}.
\ee
\label{anconflu}
\end{lem}

\begin{proof}
$A (z^q)=  \frac{(q+a) z^{q+1}}{(q+c)(1+q)}$, $A^2 \cdot c_0= c_0 \frac{(q+a)(q+a+1)}{2(q+c)(q+c+1)}z^{q+2}...$

We assume $A^m (z^q)=\frac{(q+a)_m}{(q+c)_m(q+1)_m}z^{q+m}$, where $(a)_n$ is the Pochhammer symbol. Acting $A$ on it, we get $A^{m+1} (z^q)=\frac{(q+a)_{m+1}}{(q+c)_{m+1}(q+1)_{m+1}}z^{q+m+1}$. Hence, the lemma is proved by induction.
\end{proof}

From lemma \ref{anconflu}, we see that $A$ is bounded and is a contraction operator for $|z|<1$. Putting $q=0$ in lemma \ref{anconflu} and expanding the resolvent $(1-A)^{-1}$ in a Neumann series, we obtain the regular solution :
\be
fr(z)=(1+A+A^2+...)c_0=c_0\sum_{n=1}^{\infty}\frac{(a)_n}{(c)_n}\biggl(\frac{z^n}{n!}\biggr)=\rm{_1 F_1}(a,c,z).
\ee

The contour integral representation is immediately obtained as the expression for $A^n$ is valid even when $n$ is fractional or complex. So,

\be
fr(z)= \frac{1}{2 \pi \iota}\int_{a'-\iota \infty}^{a'+\iota \infty}\Gamma(s)\Gamma(1-s)(-A)^{-s}c_0ds
= \frac{1}{2 \pi \iota}\frac{\Gamma(c)}{\Gamma(a)}c_0\int_{a'-\iota \infty}^{a'+\iota \infty}\frac{\Gamma(s)\Gamma(a-s)}{\Gamma(c-s)}(-z)^{-s}ds,
\ee
with $a'\in(0,1)$.

For $c  \notin \mathbb{R_+}$, $\int{z^{-c}dz}=\frac{z^{1-c}}{1-c}$, and therefore, for the irregular solution, we can still use lemma \ref{anconflu} with $q=1-c$. We Thus obtain
\be
fi(z)=(\sum_{n=1}^{\infty}A^n)z^{1-c}=z^{1-c}\sum_{n=1}^{\infty}\frac{(1+a-c)_n}{(2-c)_n}\biggl(\frac{z^n}{n!}\biggr)=z^{1-c}\rm{_1 F_1}(1+a-c,2-c,z).
\ee

The contour integral representation is
\be
fr(z)= \frac{1}{2 \pi \iota}z^{1-c}\frac{\Gamma(2-c)}{\Gamma(1+a-c)}\int_{a'-\iota \infty}^{a'+\iota \infty}\frac{\Gamma(s)\Gamma(1+a-c-s)}{\Gamma(2-c-s)}(-z)^{-s}ds.
\ee

\subsection{Singularities at $0$, $1$ and $\infty$}

\subsubsection{Gauss's hypergeometric function}

\begin{equation}
 (1-z)\frac{d^2\psi(z)}{dz^2}+\biggl(\frac{c}{z}-(a+b+1)\biggr)\frac{d\psi(z)}{dz} -\frac{a b}{z}\psi(z) =0.
\end{equation}

The indices at $0$ are $0$ and $1-c$.  The ODE can be written as
 \begin{equation}
 \frac{1}{z^{c}}\frac{df(z)}{dz}\biggl( z^{c} \frac{df(z)}{dz}\biggr) - \biggl((a+b+1)\frac{df(z)}{dz}+\frac{a b}{z}f(z)+ 
 z \frac{d^2\psi(z)}{dz^2}\biggr) = 0.
\end{equation}

We choose the operator $L=\int\frac{1}{z^{c} }(\int z^{c}()dz)dz$. 
The `integration constant' $f_0(z)$ for $L$, satisfying $\frac{1}{z^{c}}\frac{d}{dz}( z^{c} \frac{df_0(z)}{dz})=0$,
is $f_0(z)=c_0+\int{z^{-c}}dz$. Operating by $L$, we get $(1-A)f(z)=f_0(z)$, where
\be
\nonumber A=\int\frac{1}{z^{c}}\int z^{c}(((a+b+1)\frac{d()}{dz}+\frac{a b}{z}()+ z \frac{d^2()}{dz^2})dz)dz
\ee

Here onwards, the analysis is exactly similar to the previous (confluent hypergeometric) case. We just give the final results.
\be
A^m(z^q)=\frac{(q+a)_m(q+b)_m}{(q+c)_m(q+1)_m}z^{q+m}=\frac{\Gamma(a+m+q)\Gamma(b+m+q)\Gamma(q)\Gamma(q+c)}{\Gamma(a+q)\Gamma(b+q)\Gamma(c+m+q)\Gamma(1+m+q)}z^{q+m}.
\label{amghyp}
\ee

The regular solution obtained using (\ref{amghyp}) with $q=1-c$ in the Neumann expansion of $(1-A)^{-1}$ is
\be
fr(z)=(\sum_{i=0}^{\infty}A^n)c_0=c_0\sum_{n=1}^{\infty}\frac{(a)_n(b)_n}{(c)_n}\biggl(\frac{z^n}{n!}\biggr)=\rm{_2 F_1}(a,b,c,z).
\ee

The contour integral representation is 

\begin{align}
\nonumber fr(z)&= \frac{1}{2 \pi \iota}\int_{a'-\iota \infty}^{a'+\iota \infty}\Gamma(s)\Gamma(1-s)(-A)^{-s}c_0ds\\
&= \frac{1}{2 \pi \iota}c_0\frac{\Gamma(c)}{\Gamma(a)\Gamma(b)}\int_{a'-\iota \infty}^{a'+\iota \infty}\frac{\Gamma(s)\Gamma(a-s)\Gamma(b-s)}{\Gamma(c-s)}(-z)^{-s}ds,
\end{align}
with $a'\in(0,1)$.

For $-c \notin \mathbb{N}$, $\int{z^{-c}}=\frac{z^{1-c}}{1-c}$, and therefore the irregular solution obtained using (\ref{amghyp}) with $q=1-c$ in the Neumann expansion of $(1-A)^{-1}$ is
\be
fr(z)=(\sum_{i=0}^{\infty}A^n)c_0=c_0\sum_{n=1}^{\infty}\frac{(1+a-c)_n(1+b-c)_n}{(2-c)_n}\biggl(\frac{z^n}{n!}\biggr)=\rm{_2 F_1}(1+a-c,1+b-c,2-c,z).
\ee

The contour integral representation is 

\be
fr(z)= \frac{1}{2 \pi \iota}z^{1-c}\frac{\Gamma(2-c)}{\Gamma(1+a-c)\Gamma(1+b-c)}\int_{a'-\iota \infty}^{a'+\iota \infty}\frac{\Gamma(s)\Gamma(1+a-c-s)\Gamma(1+b-c-s)}{\Gamma(2-c-s)}(-z)^{-s}ds.
\ee

\subsubsection{The Gegenbauer polynomials}

The differential equation satisfied by the Gegenbauer polynomials, which are general cases of Legendre polynomials, is

\begin{equation}
 (z^2-1)\frac{d^2f(z)}{dz^2}+2(\beta  + 1)z\frac{df(z)}{dz}-\alpha (\alpha + 2 \beta + 1) f(z) =0.
\end{equation}

A transformation   $z\rightarrow 2 z - 1$ is needed to bring the singularity at $z = -1$ to $z = 0$.

The transformed equation is 

\begin{equation}
 z(1-z)\frac{d^2f(z)}{dz^2}+(\beta  + 1)(2z-1)\frac{df(z)}{dz} -\alpha (\alpha + 2 \beta + 1) f(z) =0.
\end{equation}

The solution to this equation follows exactly as in the case of Gauss's hypergeometric function.

\subsection*{B. Nonhomogeneous equations}

\subsection{Inhomogeneous Bessel equation}

\begin{equation}
z^2 \frac{d^2\psi(z)}{dz^2}+\frac{d\psi(z)}{dz}+(1-{\nu ^2\over{z^2}})\psi(z) =\frac{2^{1 - \nu}z^{\nu+1}}{\sqrt \pi \Gamma(\frac{1}{2} + \nu)}
\end{equation}

This inhomogeneous ODE has a particular solution which is the Struve function $H_\nu(z)$. Using equation (\ref{op_eqsoln}) we calculate the particular solution to this ODE.

\be
\psi_p(z)=z^{\nu}(1+A)^{-1}\cdot L \cdot (\frac{2^{1 - \nu}z^{-1}}{\sqrt \pi \Gamma(\frac{1}{2} + \nu)})
\ee

Using the fromulae from section (\ref{bessel})we obtain
\be
L(\frac{2^{1 - \nu}z^{-1}}{\sqrt \pi \Gamma(\frac{1}{2} + \nu)})=(\frac{2^{ - \nu}z}{\sqrt \pi \Gamma(\frac{3}{2} + \nu)})
\label{lstuve}
\ee

Using (\ref{anonxqbes}) with $q=1$ and (\ref{lstuve})

\be
A^v(L(\frac{2^{1 - \nu}z^{-1}}{\sqrt \pi \Gamma(\frac{1}{2} + \nu)}))=(-1)^v \frac{2^{-\nu}}{\Gamma(\frac{3}{2}+\nu+v)\Gamma(\frac{3}{2}+v)}\biggl(\frac{z}{2}\biggr)^{2v+1}.
\ee

The series solution and contour integral solution of $\psi_p(z)$ are therefore:

\begin{align}
\psi_p(z)=H_{\nu}(z)&=\sum_{v=0}^{\infty} \frac{(-1)^v}{\Gamma(\frac{3}{2}+\nu+v)\Gamma(\frac{3}{2}+v)}\biggl(\frac{z}{2}\biggr)^{2v+1+\nu}\\
&= \frac{1}{2 \pi \iota}\int_{a'-\iota \infty}^{a'+\iota \infty}\frac{(-1)^s\Gamma(s)\Gamma(1-s)}{\Gamma(\frac{3}{2}+\nu-s)\Gamma(\frac{3}{2}-s)}\biggl(\frac{z}{2}\biggr)^{1+\nu-2s}ds.
\end{align}

Thus we can see that complicated Green's function method to solve inhomogeneous method can be avoided using this method.

\section{Conclusion}
In this paper a new operator method is studied which is capable of giving not only analytical series solution but also exact contour integral solution to all the special functions. Calculation of complicated Green's functions is avoided in this method while solving inhomogeneous equations. Though there is a restriction $|z|<1 $ mentioned in this paper, we know that a Taylor series expansion remains valid till the nearest singularity of the differential equation and hence the solution obtained can be used well outside $|z|=1 $ if the singularity is at infinity.Only the most important equations occuring in Mathematical Physics, with 2 or atmost 3 singularities. However this method can easily solve equations with more than 3 singularity like Heun's equation much in the same manner.

 In this paper all the differential equations considered have polynomial coefficients so a operator of the form \ref{def_L} is used. In another paper this method shall be used to solve periodic differential equations with $L= \frac{1}{D^2+\omega^2}$, where D is the normal differentiation operator and $\omega$, a frequency parameter.
 
 Only analytical solutions are discussed in this paper. In another paper we shall show the numerical efficiency of the method to solve eigenvalue problems not only of 2nd order but also higher order equations. This method has also been recently generalised to solve ODEs with algebraic non linearities.The operator in that case takes the form of an infinite operator continued fraction.

{\bf Acknowledgements} I would like to thank professor Jayanta Kumar Bhattacharjee, without whose guidance this work would not have been realised. I sincerely thank Prof. Debabrata Basu for the profound influence he has had on me both professionally and others.I sincerely thank my mentor Prof. Sayan Kar for his kind support, encouragement and valuable advice. I would also like to thank Prof. S.P.Khastgir and Prof. Krishna Kumar for helpful discussions and valuable advice.

\end{document}